\documentclass{mn2e}
\usepackage{psfig}

\begin{document}

\title[BLR and emission of blazars]{The spectrum of the Broad Line
Region and the high-energy emission of powerful blazars}

\author[Tavecchio \& Ghisellini]{Fabrizio Tavecchio\thanks{E-mail:
fabrizio.tavecchio@brera.inaf.it} \& Gabriele Ghisellini\\
INAF, Osserv. Astron. di Brera, via Bianchi 46, I--23807 Merate, Italy}

\maketitle

\begin{abstract}
High-energy emission (from the X-ray through the $\gamma$-ray band) of
Flat Spectrum Radio Quasars is widely associated with the inverse
Compton (IC) scattering of ambient photons, produced either by the
accretion disk or by the Broad Line Region, by high-energy electrons
in a relativistic jet. In the modelling of the IC spectrum one usually
adopts a simple black body approximation for the external radiation
field, though the real shape is probably more complex. The knowledge
of the detailed spectrum of the external radiation field would allow
to better characterize the soft-medium X-ray IC spectrum, which is
crucial to address several issues related to the study of these
sources. Here we present a first step in this direction, calculating
the IC spectra expected by considering a realistic spectrum for the
external radiation energy density produced by the BLR, as calculated
with the photoionization code CLOUDY. We find that, under a wide range
of the physical parameters characterizing the BLR clouds, the IC
spectrum calculated with the black-body approximation reproduces quite
well the exact spectrum for energies above few keV. In the soft energy
band, instead, the IC emission calculated using the BLR emission shows
a complex shape, with a moderate excess with respect to the
approximate spectrum, which becomes more important for decreasing
values of the peak frequency of the photoionizing continuum. We also
show that the high-energy spectrum shows a marked steepening, due to
the energy dependence of the scattering cross section, above a
characteristic energy of 10-20 GeV, quasi independent on the Lorentz
factor of the jet.
\end{abstract}


\section{Introduction}

Powerful gamma-ray emission (at energies $E>100$ MeV) is a distinctive
feature of the group of Active Galactic Nuclei collectively known as
{\it blazars}. As clearly shown by their Spectral Energy Distributions
(e.g. Fossati et al. 1998), the gamma-ray emission belongs to a
component different from the low-energy synchrotron bump, commonly
ascribed to relativistic electrons in a jet pointing close to the
observer. Among the models advanced to account for the high-energy
component, the most widely discussed (also known as ``leptonic
models'' to distinguish them from the other models involving hadronic
reactions, e.g. M{\"u}cke et al. 2003) assumes that the emission is
produced through the inverse Compton scattering between the
synchrotron-emitting electrons and either the synchrotron photons (in
the {\it Synchrotron Self-Compton} model, Maraschi et al. 1992) or
photons from the regions outside the jet (the so-called {\it External
Compton} models). In turn, external radiation can be dominated by the
emission of the accretion disk (Dermer \& Schlickeiser 1993),
reaching the jet either directly or through reflection by free
electrons, or by the emission of the clouds belonging to the Broad
Line Region (Sikora et al. 1994). The most powerful blazars (Flat
Spectrum Radio Quasars) display luminous broad emission lines and, in
some cases, also show the clear presence of the ``blue bump'', usually
associated with the direct emission from the accretion disk (Sun \&
Malkan 1989). These indications strongly suggest that in FSRQs the
high-energy component is dominated by the radiation produced through
the EC process, likely with photons produced by the BLR (although the
SSC emission can provide an important contribution in the medium-soft
X-ray band).

Current models of the EC emission adopts a quite simplified
description of the external radiation field, usually approximated with
a bumpy, black-body like, spectrum (e.g. Inoue \& Takahara 1996,
Dermer, Sturner \& Schlickeiser 1997, Ghisellini et al. 1998,
B{\"o}ttcher \& Bloom 2000). In the case of EC models considering BLR
radiation, this approach is justified with the argument that the
spectrum of the seed photons is dominated by few strong emission
lines (in particular the Ly$_{\alpha}$ line). However, the actual
spectrum of the soft photons can be more complex: besides the lines,
there is the important contribution of the continuum emission, which
(depending on the parameters characterizing the BLR clouds) could
imprint important signatures in the EC emission. In particular, the
shape of the soft radiation field would be reflected in the EC
spectrum in the medium-soft X-ray band, produced by the electrons at
the lower end of the energy distribution. The determination of the
exact spectral shape in this band would be crucial for several
reasons: 1) interesting constraints on the bulk Lorentz factor of the
jet and the minimum electron energy can be derived from the observed
spectral hardening below about 1 keV, expected to trace the shape of
the seed photons (e.g. Tavecchio et al. 2007; Ghisellini et
al. 2007); 2) alternatively, if the observed hardening below 1 keV in
some high-redshift blazars is attributed to the presence of absorbing
material at the quasar redshift (e.g. Elvis et al. 1994, Cappi et
al. 1997, Reeves et al. 1997, Fiore et al. 1998, Fabian et
al. 2001a,b, Bassett et al. 2004, Yuan et al. 2005, Page et al. 2005),
the knowledge of the underlying continuum is essential to characterize
the amount and the physical state of the intervening gas; 3) a
correct description is also essential in view of the possible
contribution in this band of the bulk-Comptonized radiation from cold
leptons in the jet (Begelman \& Sikora 1987, Celotti, Ghisellini \&
Fabian 2007).

Apart from these aspects, specifically related to the modelling of the
soft X-ray emission, the knowledge of the diffuse radiation filling
the regions surrounding the jet has a strong impact in the
determination of the opacity of the environment to the gamma-ray
photons (Liu \& Bai 2006, Reimer 2007). The assessment of the effects
of this ``internal absorption'' is mandatory in view of the
possibility to use the absorption of gamma-rays to probe the
optical-UV cosmic background (Chen et al. 2004, Reimer 2007).

With these motivations, in this work we intend to explore some of the
implications of the detailed modelling of the diffuse radiation field
from the BLR for the expected EC spectrum of powerful blazars. In
particular, we calculate the spectrum emitted by the clouds, for
a relatively large range of the parameters, using the photoionization
code CLOUDY (Section 2). These spectra are then used to calculate the
EC spectra (Section 3), which can be compared with the spectra
obtained using the black-body approximation. Finally, in Section 4 we
discuss the results.

\section{The external radiation field}

\subsection{Setting the stage}

The geometry assumed for the modelling is depicted in Fig. 1. The
central accretion flow illuminates the clouds residing in the BLR
(characterized by the [total: ionized $+$ neutral] hydrogen density
$n$ and the hydrogen column density $N_H$), assumed to be a spherical
shell with inner radius $R_{\rm in}$ and thickness $\Delta R$. In the
calculation we assume that the clouds cover a fraction
$C=\Omega/4\pi=0.1$ of the solid angle viewed from the central
illuminating source. The emission from the illuminated face of the
clouds (comprising the reflected incident continuum, the emission
lines and the diffuse continuum from clouds, see e.g. Korista \&
Ferland 1998) is assumed to isotropically fill the region within
$R_{\rm in}$ and it is calculated with version 05.07 of CLOUDY,
described by Ferland et al. (1998)\footnote{See also {\tt
http://www.nublado.org/}}. For simplicity we  discuss only the
case of solar abundance and we neglect the effect on the clouds of the
net diffuse emission from the clouds themselves. We assume that the
distance of the jet from the central black hole is negligible compared
with the radius of the BLR. This assumption greatly simplifies the
geometrical treatment. Furthermore we do not consider the possible
contribution of photons originating in the disk, reaching the jet
either directly or after being re-isotropized by intercloud gas
(e.g. Celotti et al. 2007).

\begin{figure}
\centerline{ \psfig{figure=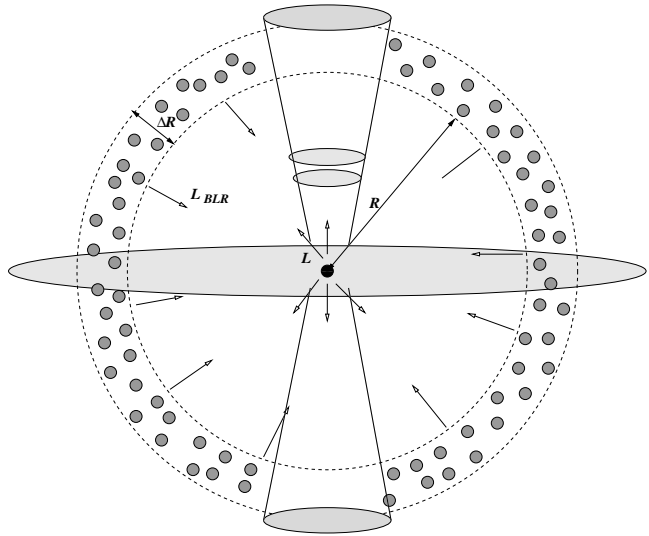,width=8.5cm}}
\caption{Sketch of the geometry assumed in the model (not to
scale). The (uniform) BLR is assumed to be a spherical shell with
thickness $\Delta R$ and inner radius $R_{\rm in}$, illuminated by the
central continuum with luminosity $L_{\rm ill}$. See text for
details.}
\end{figure}

We adopt the spectrum of the illuminating continuum modelled as a
combination of a UV bump with an X-ray power-law $\nu ^{-1}$ usually
assumed in this kind of calculations (AGN model in CLOUDY,
e.g. Korista \& Goad 2001 and references therein):
\begin{equation}
L_{\rm ill}(\nu)\propto \nu^{-0.5} \exp\left(-\frac{h\nu}{kT_{BB}}\right)
 \left[ 1- \exp\left(-\frac{h\nu}{kT_{IR}}\right) \right]+A\nu^{-1}
\end{equation}
\noindent
Above 100 keV we assume a steep ($\nu^{-2}$) power law mimicking the
expected cut-off. This shape reproduces reasonably well the average
spectra measured from AGNs (Francis et al. 1993; Elvis et
al. 1994). The bump is assumed to have a cut-off in the IR band at 9.1
$\mu$m (Korista et al. 1997), since most of the IR emission from AGNs
is believed to originate in warm dust outside the BLR. The
normalization constant $A$ is fixed to provide a value of $\alpha
_{ox}=1.4$, as typically observed in AGNs (e.g. Zamorani et
al. 1981). The continuum is then fixed by assuming a normalization for
the spectrum (parametrized using the {\it bolometric} luminosity
$L_{\rm ill}$) and the temperature of the bump, $T_{BB}$.

Using this model for the continuum we calculate the diffuse emission
of the BLR for the hydrogen density and the hydrogen column density of
the clouds in the ranges $n=10^9-10^{11}$ cm$^{-3}$ and
$N_H=10^{22}-10^{24}$ cm$^{-2}$, respectively. These intervals cover
the region of the parameter space allowed by detailed modelling of the
lines observed for few selected (radio-quiet) AGNs (e.g. Korista \&
Goad 2000; Kaspi \& Netzer 1999). A {\it caveat} (which applies also
to the choice of the spectral shape of the illuminating continuum) is
that the present knowledge of the geometry and the physical state of
the gas in the BLR mainly relies on the studies of (few) well-observed
radio-quiet AGNs (mainly Seyfert galaxies). Indeed it is conceivable
that for the sources for which we are interested here, namely
powerful, radio-loud quasars, the conditions could be different. Note
moreover that, for simplicity, we model the BLR with a unique value of
the physical parameters, although the recent detailed models of
emission lines quoted above seem to support the presence of a
stratified BLR, with the parameters (density, column density, covering
factor) varying with distance from the central source.

\subsection{The spectrum of the BLR}

In Fig. 2 we report some examples of the spectra of the diffuse BLR
emission [plotted in the $\nu L(\nu)$ representation], calculated for
different values of $n$ and $N_H$, assuming the values $T_{BB}=10^5$ K
(which fixes the peak of the blue bump around $\sim 10^{15}$ Hz,
dashed lines), $L_{\rm ill}=3\times 10^{47}$ erg/s and $R_{\rm
in}=10^{18}$ cm.

\begin{figure*}
\hskip 0.5 truecm \psfig{figure=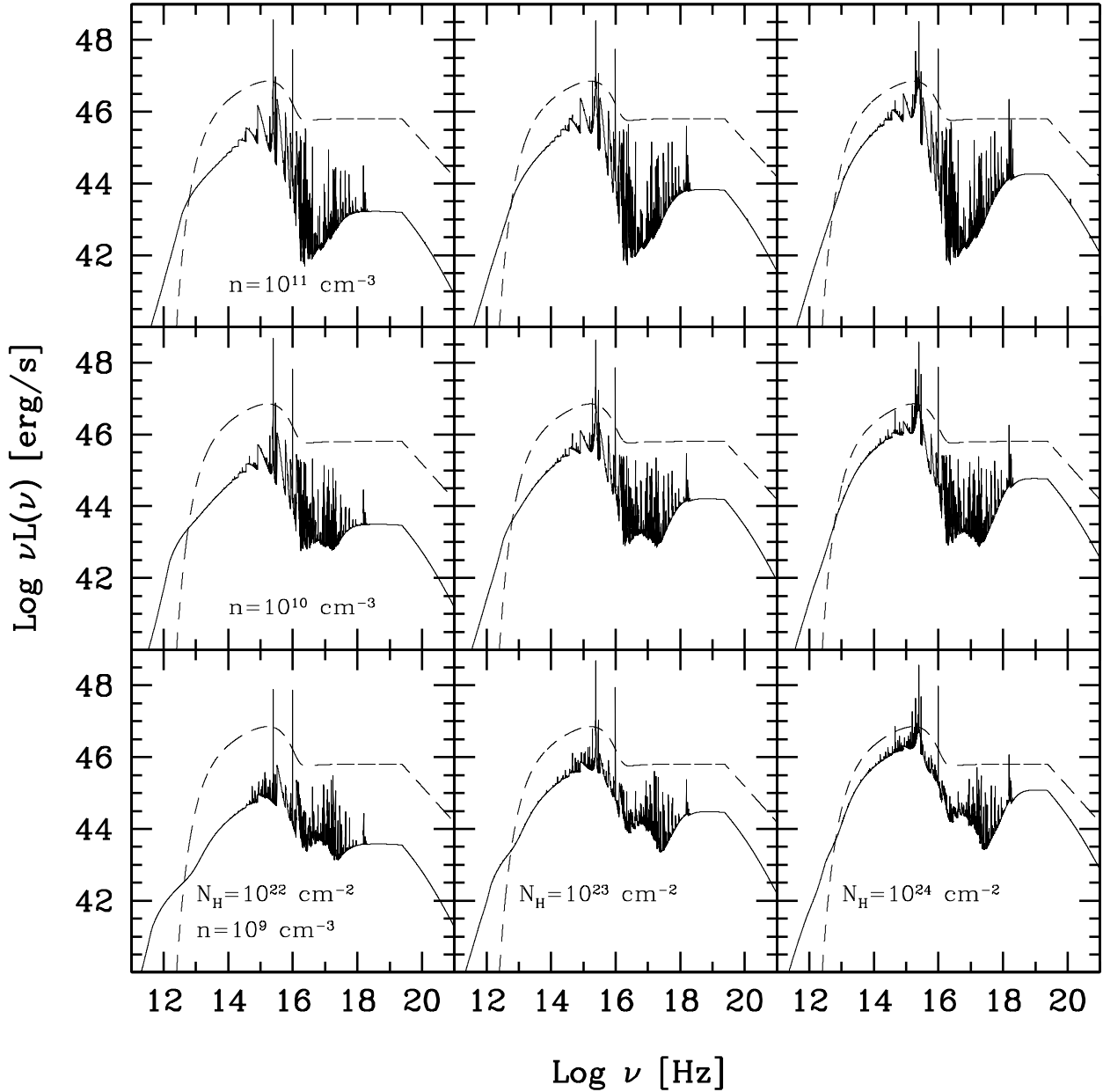,angle=0,width=20cm}
\caption{Examples of the spectral energy distribution of the emission
from the Broad Line Region calculated for different values of gas
density, from $n=10^9$ cm$^{-3}$ (bottom) to $n=10^{11}$ cm$^{-3}$
(top), and column density, from $N_H=10^{22}$ cm$^{-2}$ (left) to
$N_H=10^{24}$ cm$^{-2}$ (right). The SEDs indicated with dashed lines
show the illuminating continuum, for which $T_{BB}=10^5$ K. The model
assumes a total luminosity of the incident continuum of $L_{\rm
ill}=3\times 10^{47}$ erg/s and the BLR clouds (with covering factor
$C=0.1$) are assumed to reside in a spherical shell with thickness
$\Delta R=4\times 10^{18}$ and inner radius $R_{\rm in}=10^{18}$
cm.}
\end{figure*}

The large value of the luminosity is suitable to characterize the
powerful quasars at medium-high redshift. Disk luminosity $L> 10^{47}$ erg/s
have been inferred from the optical-UV data of sources recently
observed with {\it Swift/UVOT} (e.g. Tavecchio et al. 2007, Sambruna
et al. 2007). The radius of the BLR has been fixed to a value between
those provided by the relations found by Kaspi et al. (2007) and Bentz
et al. (2006). These relations connect the luminosity at a given
wavelength to the distance of the clouds emitting a given line,
estimated through the reverberation mapping technique. The relations
recently derived by Kaspi et al. is based on the CIV line and the
continuum is evaluated at $\lambda = 1350$ \AA, while the Bentz et
al. relation uses the H$\beta$ line and the continuum at $\lambda =
5100$ \AA. In both cases the relation has the same dependence,
$R_{BLR}\propto (\lambda L_{\lambda})^{0.5}$. The values of the
monochromatic luminosities $L_{1350}$ and $L_{5100}$ depend on the
spectral shape of the illuminating continuum. For the shape and the
luminosity assumed here, the two relations provide relatively
different values for $R_{BLR}$, $R_{BLR}=5\times 10^{17}$ cm and
$R_{BLR}=3\times 10^{18}$ cm for Kaspi et al. and Bentz et al.,
respectively. For the calculation shown in Fig. 2 we assume the
(logarithmic) average value $R_{\rm in} = 10^{18}$ cm.

It is worth noting that, although calculated for a specific set of
parameters, the spectra reported in Fig. 2 can be easily generalized
to different values of $L_{\rm ill}$ and BLR radius. Indeed, the
normalization of the diffuse spectrum simply scales as the luminosity
$L_{\rm ill}$, while the shape (continuum and lines) only depends (at
the first order) on the ionization parameter $\xi=L_{\rm ill}/nR_{\rm
in}^2$ (in terms of $\xi$, the spectra in Fig. 2 correspond to
$\xi=300, 30$ and 3, from bottom to top, respectively). Different
combination of $L_{\rm ill}$, $R_{\rm in}$ and $n$ can thus produce
similar spectra. Note that, if $L_{\rm ill}$ and $R_{\rm in}$ are
linked by a relation of the form $R_{\rm in}\propto L_{\rm ill}^{0.5}$
as discussed above, the ionization parameter in different sources
would only depend on the density of the clouds (assuming similar
spectra for the photoionizing continuum), since the ratio $L_{\rm
ill}/R_{\rm in}^2$ would be a fix number in different sources.

As expected, the total luminosity of the diffuse emission is dominated
by the optical-UV emission, and, in particular, by few prominent lines
in this band, most notably the Ly$_{\alpha }$ lines of H (1216 \AA)
and HeII (303 \AA). In all cases the underlying continuum around the
peak (dominated by recombinations and scattering, both Thomson and
Rayleigh) contributes to few percent of the total luminosity. However,
the continuum (mainly due to free-free emission) dominates below the
peak (below $\sim 10^{15}$ Hz), making the spectrum much broader than
the black body approximation. As we discuss later, this fact has
a direct consequence on the derived EC spectrum.

Some trends are clearly visible in Fig. 2. Increasing $N_H$ (left to
right) has the effect to increase the total emitted luminosity and,
importantly, to increase the relative contribution of the optical-UV
continuum with respect to the emission lines. The increase of the
density $n$ (bottom to top), instead, has the effect of reducing the
ionization parameter $\xi $ and leads to increase the importance of
the recombinations (as witnessed by the prominent edges).

An interesting feature of these spectra is the presence of an
important X-ray component, originating from the Compton reflection on
the cloud gas (analogously to the ``reflection bump'' from accretion
disks). As expected, the fraction of the X-ray continuum (and of the
overall continuum) reflected by the BLR increases (from left to right)
with the column density $N_H$. Increasing (from bottom to top) the
density $n$, instead, has the effect of reducing the reflected X-ray
luminosity, since the ionization parameter $\xi $ (and thus, for the
same value of the column density, the number of free electrons
available for scattering) decreases. The decreased ionization state of
the gas is also responsible for the deepening of the depression
visible in the soft X-ray band, caused by photoelectric
absorption. The presence of an important diffuse X-ray emission inside
the BLR could have important consequences for the propagation of
$\gamma $-rays with $E\sim 100$ MeV produced by the jet (Ghisellini et
al. 2008, in preparation).

\begin{figure}
\centerline{\psfig{figure=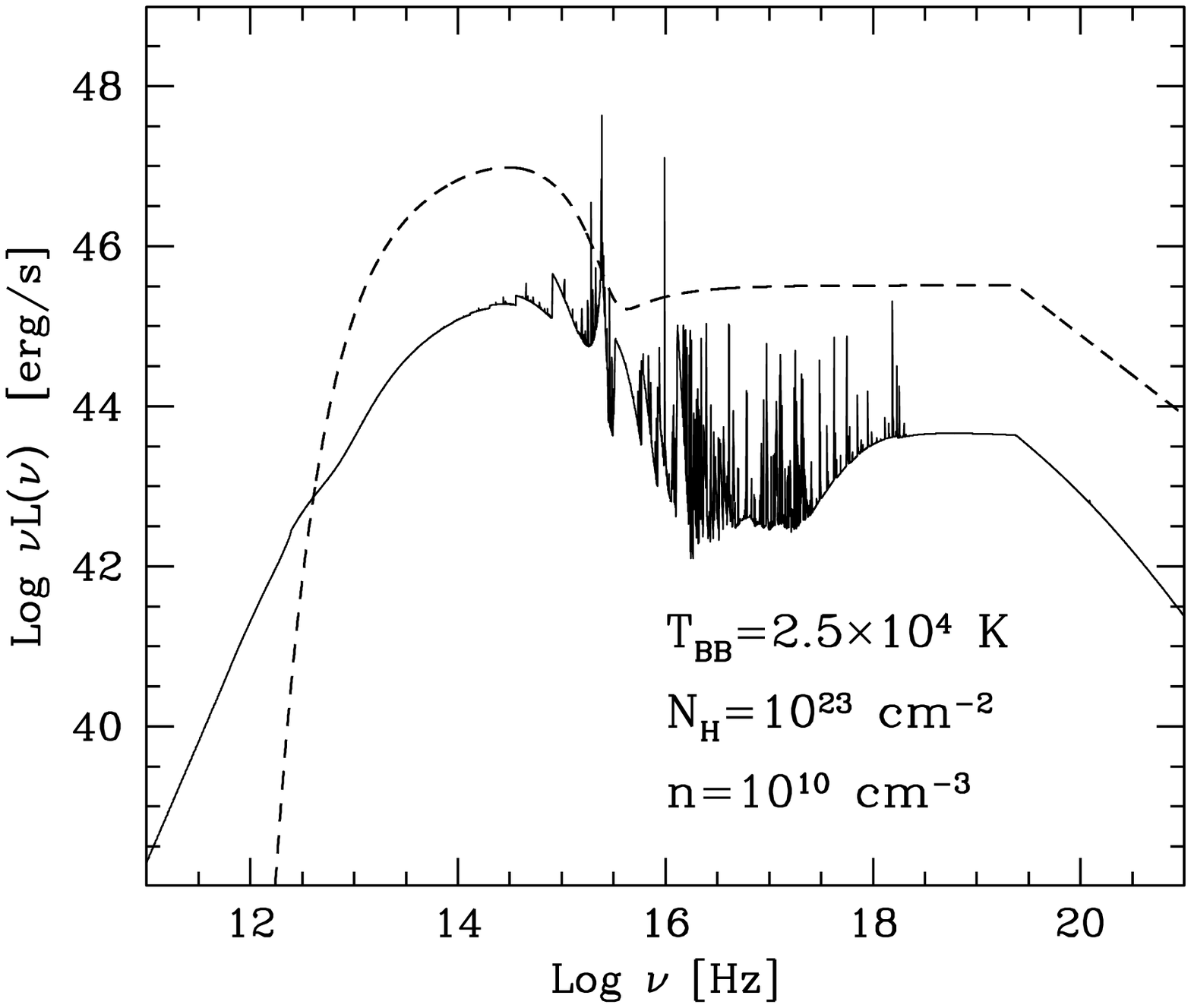,angle=0,width=8.5cm}}
\caption{SED of the BLR emission (and the corresponding illuminating
continuum, dashed line) calculated for a temperature of the bump,
$T_{BB}=2.5\times 10^4$ K, density $n=10^{10}$ cm$^{-3}$, column
density $N_H=10^{23}$ cm$^{-2}$ and the other parameters fixed to the
values used in Fig. 2. Note the more prominent continuum with respect
to the higher-temperature case shown in Fig. 2.}
\end{figure}

The basic features of the BLR emission do not strongly depend on the
actual shape of the photoionizing continuum. The effect of a different
temperature $T_{BB}$ is illustrated by Fig. 3, reporting the SEDs of
the BLR reflection calculated for the case $n=10^{10}$ cm$^{-3}$,
$N_H=10^{23}$ cm$^{-2}$ and $T_{BB}=2.5\times 10^4$ K. The overall
shape of the SEDs is similar to the high-temperature case (Fig. 2,
central panel). However, the lower temperature has the effect to
decrease the relative flux of the ionizing photons, leading to
decrease the luminosity of the emission lines relative to that of the
continuum (whose peak, moreover, moves to lower frequencies by a
factor of $\approx 2$). As we show later, the increasing
importance of the continuum with respect to the lines can
significantly affect the resulting EC spectrum.

\begin{figure}
\centerline{\psfig{figure=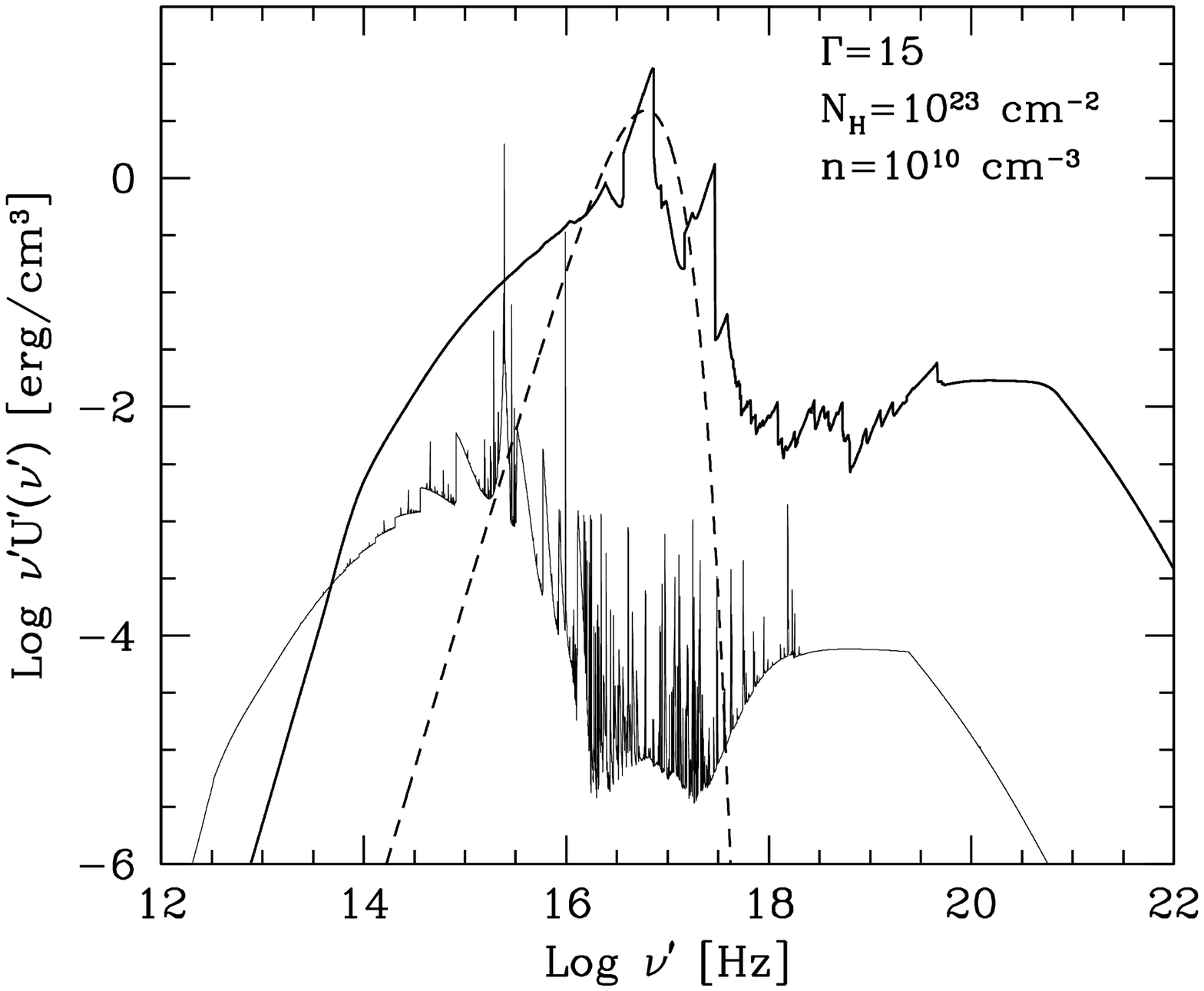,angle=0,width=8.5cm}}
\caption{Radiation energy density of the BLR emission (calculated for
$n=10^{10}$ cm$^{-3}$, column density $N_H=10^{23}$ cm$^{-2}$ and the
other parameters fixed to the values used in Fig. 2) transformed into
the jet frame. The bulk Lorentz factor of the jet is fixed to $\Gamma
=15$. For comparison we report the corresponding energy density in the
quasar frame, before the boosting (in this case both the frequency and
the energy density are measured in the quasar rest frame). The dashed
line shows the black-body spectrum used to calculate the approximate
EC emission.}
\end{figure}

\section{Inverse Compton spectra}

\subsection{The external energy density in the jet frame}

To calculate the IC emission produced by relativistic electrons
scattering off the diffuse emission of the BLR we have first to
transform the spectrum of the seed photons to the jet frame, assumed
to move with bulk Lorentz factor $\Gamma $. We consider only the
photons emitted by the emispheric portion of the BLR above the
accretion disk, since the disk ``occults'' the other emisphere.

Because of relativistic aberration, all photons are observed in the
jet frame as coming into a narrow cone of semiaperture $\sim 1/\Gamma$
with the axis aligned along the jet speed. Within this cone, the
radiation field coming from different angles are observed to have
different frequencies and intensities, maximum along the axis and
minimum at the edge. The result is that an external monochromatic
radiation field with frequency $\nu _o$ is observed in the jet frame
to have a hard spectrum ($\propto \nu^{\prime 2}$, we use the prime to
indicate quantities evaluated in the jet frame), between
$\nu_1^{\prime}=\Gamma \nu _o $ (photons coming from the cone edge)
and $\nu_2^{\prime}\simeq 2\Gamma \nu _o$ (photons at the axis of the
cone). This can be shown in the following way. For an external
radiation field with specific intensity $I(\nu)$, the corresponding
radiation energy density in the jet frame is:

\begin{equation}
U^{\prime}(\nu^{\prime}) = \frac{1}{c} \int I^{\prime}(\nu^{\prime})
d\Omega^{\prime} = \frac{2\pi}{c} \int _0 ^{\beta} I(\nu)
\left(\frac{\nu^{\prime}}{\nu}\right)^3 d\mu^{\prime}
\end{equation}
\noindent
where we use the invariance of the ratio $I(\nu)/\nu^3$ (e.g. Rybicki
\& Lightman 1979) and the integral is extended over the angles $\theta
^{\prime}\equiv \arccos \mu ^{\prime}$ in the range $0-1/\Gamma$. Now
we change the variable of integration from $\mu ^{\prime}$ to $\nu$
using:
\begin{equation}
\nu^{\prime }=\frac{\nu}{\Gamma (1-\beta \mu^{\prime})} \rightarrow
\mu ^{\prime}=\frac{\Gamma \nu^{\prime}-\nu }{\Gamma \beta \nu
^{\prime}} \rightarrow d\mu^{\prime} = - \frac{d\nu}{\Gamma \beta \nu^{\prime}}
\end{equation}
\noindent
and we obtain:
\begin{equation}
U^{\prime}(\nu^{\prime}) = \frac{2\pi}{\Gamma \beta c} \nu ^{\prime 2}
\int _{\nu_1} ^{\nu_2} \frac{I(\nu)}{\nu^3} d\nu
\label{uin}
\end{equation}
\noindent
where the new limits become $\nu_1=\nu^{\prime}/[\Gamma(1+\beta)]$ and
$\nu_2=\nu^{\prime}/\Gamma$.  In the case of a monochromatic external
field, $I(\nu)=I_o\delta(\nu-\nu_o)$ we obtain:
\begin{equation}
U^{\prime}(\nu^{\prime}) = \frac{2\pi I_o}{\Gamma \beta c \nu_o^3}\, \nu
^{\prime 2} \,\,\,\,\,\, {\rm for} \,\,\,
\Gamma\nu_o < \nu^{\prime} < (1+\beta)\Gamma \nu_o
\end{equation}
\noindent
One can check that, es expected, $U^{\prime}=\int U(\nu^{\prime})
d\nu^{\prime} = (7/3)\, \Gamma ^2 \int U(\nu) d\nu \sim \Gamma ^2 U$.

Using Eq. (\ref{uin}) we can now calculate the energy density of the
BLR radiation in the jet frame. An example is reported in Fig. 4 for
the reference case $n=10^{10}$ cm$^{-3}$, $N_H=10^{23}$ cm$^{-2}$,
$T_{BB}=10^5$ K. The effect of the boosting is clearly visible for the
Ly$_{\alpha}$ line, which is transformed into a broad (extending from
$\nu^{\prime} =\Gamma \nu _{Ly_{\alpha}}$ to $\nu^{\prime} \simeq
2\Gamma \nu _{Ly_{\alpha}}$), hard ($\propto \nu^{\prime 2}$)
feature. Due to the relativistic broadening most of the lines merge in
few broad features. In view of the comparison between the EC spectrum
calculated by using the BLR emission and the approximated one, we
define the black body which best approximate the boosted BLR emission,
with the peak corresponding to the frequency $\nu ^{\prime} _{BB}=1.5
\Gamma \nu _{Ly_{\alpha} }$ and with normalization equal to that of the
integral of the boosted BLR radiation energy density (dominated by the
${Ly _{\alpha} }$ line). It is shown in Fig. 4 by the dashed line.

\begin{figure*}
\hskip 0.5 truecm
\psfig{figure=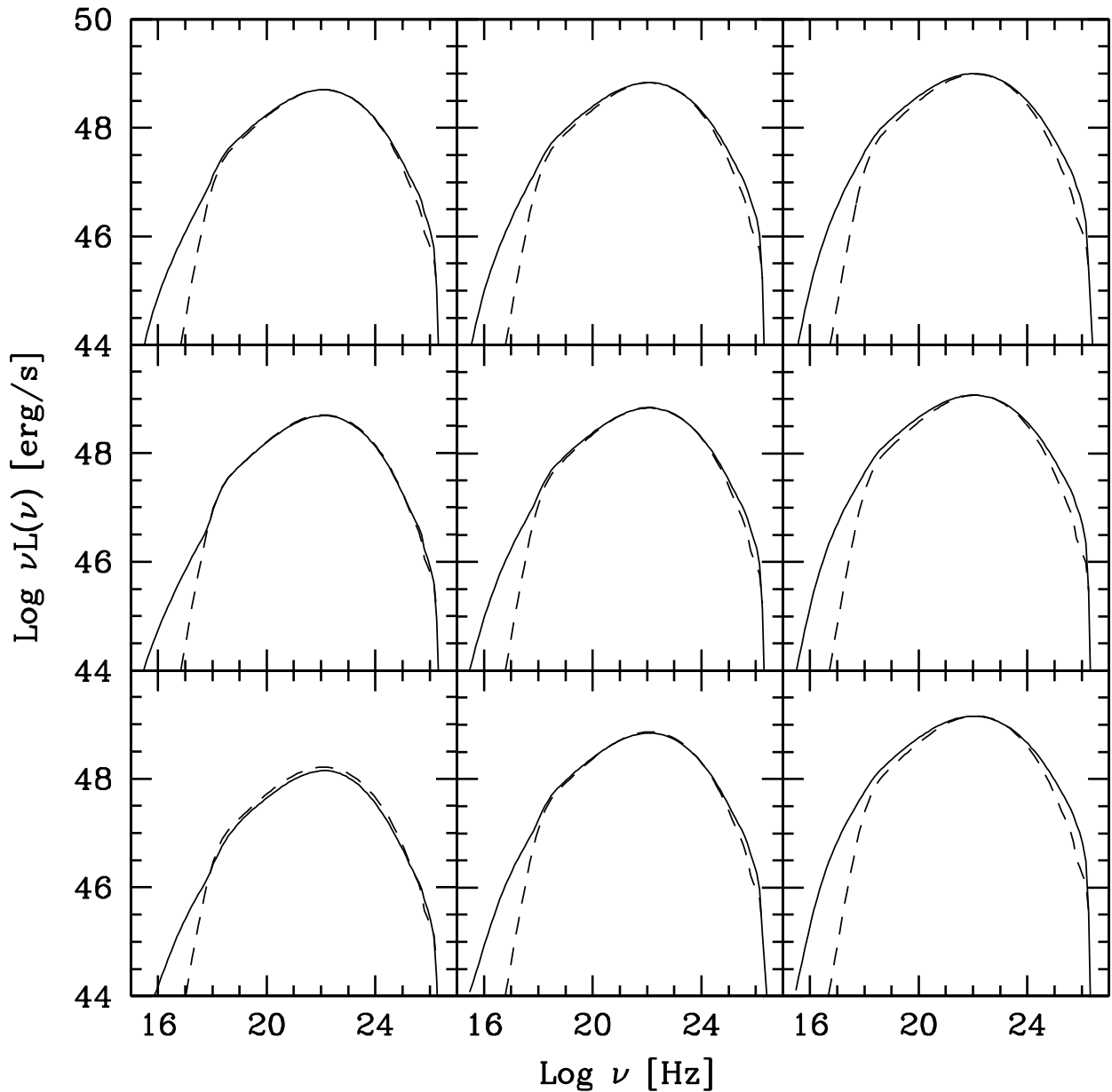,angle=0,width=20cm}
\caption{EC spectra calculated for the BLR energy densities reported
in Fig. 2 ({\it EC/BLR}, solid lines) and with the approximating
black-body ({\it EC/BB}, dashed lines). Clearly, the two kind of
spectra agree well for frequencies above $\sim 10^{18}$ Hz. Below this
frequency, the EC/BB spectra show a continuum harder than the
corresponding EC/BLR spectra. The discrepancy increases for increasing
values of density and column density of the clouds.}
\end{figure*}

\subsection{The EC spectrum}

Using the energy density in the jet frame we can now compute the EC
spectrum. To this aim we use the radiative model fully described in
Maraschi \& Tavecchio (2003). Briefly, we assume that the emission
region, moving with bulk Lorentz factor $\Gamma$ at an angle $\theta
_w=1/\Gamma $ (therefore the relativistic Doppler
factor\footnote{given by $\delta=[\Gamma(1-\beta\cos \theta
_w)]^{-1}$} is $\delta=\Gamma$) is isotropically filled with
relativistic electrons, assumed to have a smoothed broken power-law
electron energy distribution, extending from the energy $\gamma _{\rm
min}mc^2$ to $\gamma _{\rm max}mc^2$, with slopes $n_1$ and $n_2$
below and above the break located at Lorentz factor $\gamma _{\rm
p}$. This (purely phenomenological) form has been assumed to reproduce
the observed shape of the blazar SEDs, without any specific assumption
on the acceleration/cooling mechanism acting on the particles.

In Fig. 5 we report the spectra that we obtain for all the BLR models
shown in Fig. 2 ({\it EC/BLR}), using $\gamma _{\rm min }=1$,$\gamma
_{\rm p}=100$, $\gamma _{\rm max}=10^4$, $n_1=2$, $n_2=3.5$, $\Gamma
=15$ (values usually derived from the modelling of FSRQs, see
e.g. Ghisellini et al. 1998; Maraschi \& Tavecchio 2003). The electron
energy distribution has been normalized arbitrarily. For comparison,
we also report (dashed line) the spectra calculated using the
black-body approximation defined above ({\it EC/BB}). This can be
simply described by three power laws with smooth transitions: 1) an
extremely hard power law ($\propto \nu^2$) below a ``break frequency''
$\nu _{\rm br}\simeq \nu ^{\prime }_{BB} \Gamma \gamma _{\rm min }^2\sim
10^{18}$ Hz
; 2) a power law with spectral index $\alpha _1=(n_1-1)/2$ between
$\nu _{\rm br}$ and $\nu _{\rm p}\simeq \nu ^{\prime} _{BB} \Gamma
\gamma _{\rm p}^2$ (marking the position of the EC peak); 3) a steeper
power law with slope $\alpha _2=(n_2-1)/2$ from $\nu _{\rm p}$ up to
the frequency $\nu _{\rm max}$.

\subsubsection{The low-energy spectrum}

In all cases the {\it EC/BLR} spectra agree quite well with the
{\it EC/BB} ones down to the frequency $\nu _{\rm br }$, marking the
position of the peak of the emission from the electrons at the lower
end $\gamma _{\rm min}$. 
Below this frequency, the slope reflects the shape of the spectrum of
the seed photons. Thus, the flux of the {\it EC/BB} decreases as
$\nu^2$. Instead, the corresponding {\it EC/BLR} spectra shows a tail
in excess of the {\it EC/BB} spectra, clearly following the excess of
soft photons with respect to a black-body shown by the BLR continuum
(see Fig. 4). In other words, soft X-ray spectra calculated with the
{\it EC/BLR} tends to be softer than the corresponding {\it EC/BB}
spectra. To better clarify this point we report in Fig. 6 the
{\it EC/BB} (top) and the {\it EC/BLR} (bottom) spectra (calculated
for $n=10^{10}$ cm$^{-3}$ and $N_H=10^{23}$ cm$^{-2}$) with the
contribution of electrons in different ``slices'' of energy. For
sufficiently thin slices the EC spectral shape simply reflects the
incoming spectrum of the seed photons (broadened by the boosting into
the jet frame). The broad BLR spectrum results in a tail (roughly
hardening from $\nu^0$ to $\nu^2$) in excess to the narrow peaked
black-body shape.

\begin{figure}
\centerline{\psfig{figure=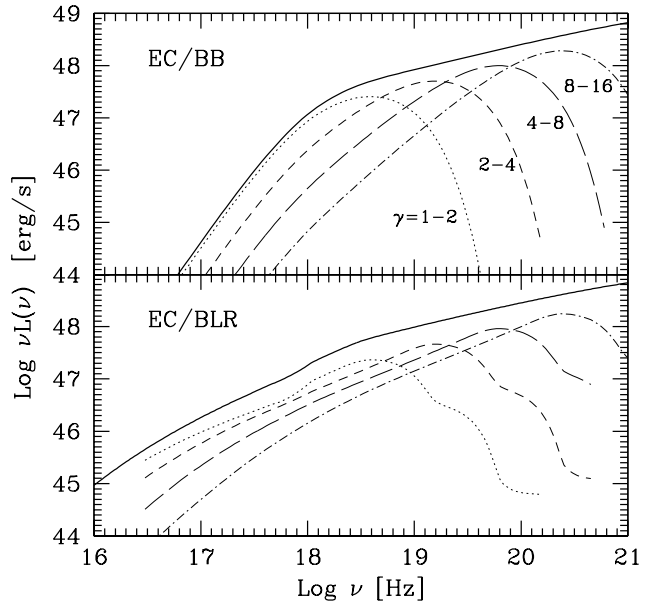,angle=0,width=9.5cm}}
\caption{{\it EC/BB} (top panel) and {\it EC/BLR} (bottom panel)
spectra (solid lines, for the case $n=10^{10}$ cm$^{-3}$ and
$N_H=10^{23}$ cm$^{-2}$) together with the contribution of electrons
in small ``slices'' of energy, $\gamma=$1--2, 2--4, 4--8 and
8--16. For sufficiently small ranges in energy these components simply
reflect the underlying shape of the seed photon distribution. The BLR
spectrum, broader than a black-body, is the reason of the soft tail
displayed by the {\it EC/BLR} spectrum.}
\end{figure}

{\it EC/BLR} spectra in Fig. 5 shows clear trends related to the
features of the corresponding BLR spectra. In particular, the
luminosity increases for increasing $N_H$ (left to right), following
the corresponding increase of the luminosity displayed by the BLR
emission. Moreover, a better agreement between the {\it EC/BLR} and
{\it EC/BB} in the soft X-ray band spectra is visible for low
$N_H$. This is clearly related to the fact that low column densities
imply a minor contribution of the continuum to the total BLR
luminosity. At large column densities, instead, the relative
importance of the continuum grows and therefore the discrepancy
between the two kind of spectra becomes more and more important. Note
that in this case even the power law above 1 keV becomes slightly
softer.

\begin{figure}
\centerline{\psfig{figure=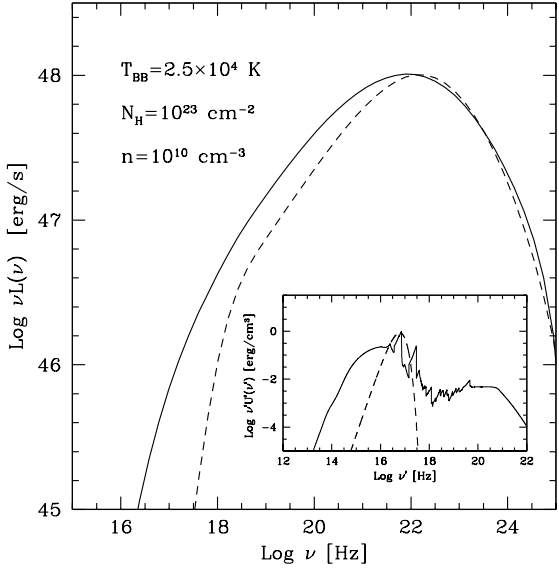,angle=0,width=8.5cm}}
\caption{EC spectrum calculated for the reference value of the BLR
  parameters and $T_{BB}=2.5\times 10^4$ K. The dashed line reports
  the corresponding EC spectrum calculated with the black-body
  approximation. The insert shows the BLR radiation energy density (in
  the $\nu U(\nu)$ representation) obtained in this case, together
  with the corresponding black body (dashed line). The clear
  discrepancy between the EC spectra calculated using the two shapes
  for the radiation energy density is the effect of the minor prominent
  ``shoulder'' displayed by the BLR spectrum below the Ly$_ \alpha $
  line.}
\end{figure}

All the effects described above become more important in the case of
the low-temperature disk, reported in Fig. 7. In this case, as already
noted, the continuum/line luminosity ratio is larger than in the
high-temperature case (as clearly visible in the spectrum of the
boosted BLR radiation energy density shown in the inset) and thus the
discrepancy between the {\it EC/BB} and the {\it EC/BLR} increases. In
particular, the {\it EC/BLR} continuum appears much broader than the
corresponding {\it EC/BB} spectrum.

\subsubsection{The high-energy spectrum}

At high frequencies both the {\it EC/BLR} and the {\it EC/BB} spectra
shows a rapid decrease above $\sim 10^{24-25}$ Hz. The reason for such
a cut-off in the spectrum is the decreased rate of the IC scatterings
in the Klein-Nishina (KN) regime. To derive an approximate estimate of
the cut-off frequency expected from the KN, $\nu _{\rm KN}$, let us
assume a monochromatic radiation field with frequency $\nu _o$. In the
rest frame of the jet the boosted frequency is $\nu ^{\prime}_o\simeq
2\Gamma \nu _o$. Electrons with Lorentz factors $\gamma $ scatter
these photons in the Thomson regime as long as $\gamma h \nu
^{\prime}_o < (3/4) m_e c^2$, producing photons with frequency $\nu
^{\prime }_C\simeq (3/4) \nu ^{\prime}_o \gamma ^2$. Using this
expression for the frequency of the outcoming photon, the condition
that the scattering happens in the Thomson limit can be written as
$\nu ^{\prime} _C < 3(m_ec^2/h)^2 / 4\nu ^{\prime}_o$. These photons
are observed at the frequency $\nu _C\simeq \delta \nu ^{\prime}
_C$. Thus, using quantities in the observer frame, we can express the
condition above as:

\begin{eqnarray}
\nu _C < \nu _{\rm KN} &\equiv& \frac{3\delta}{8\Gamma} \left(
\frac{m_e c^2}{h} \right)^2 \frac{1}{\nu _o} \nonumber \\
&=& 5.6\times 10^{24} \nu
_{o,15}^{-1} \frac{\delta}{\Gamma}\,\, {\rm Hz} = \frac{22.5}{\nu _{o,15}} 
\frac{\delta}{\Gamma} \,\, {\rm GeV}
\label{nukn}
\end{eqnarray}

\begin{figure}
\centerline{\psfig{figure=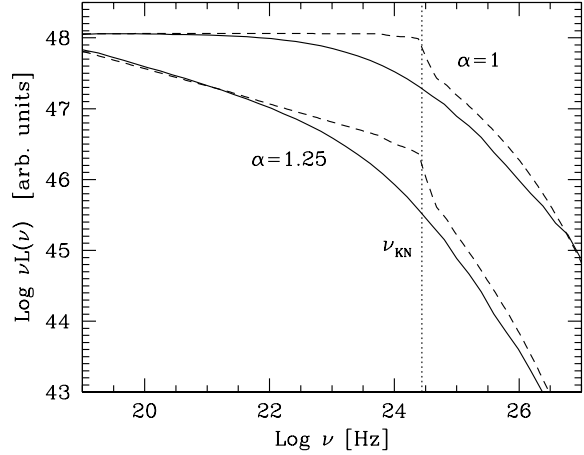,angle=0,width=8.5cm}}
\caption{{\it EC/BLR} spectra calculated assuming $n=10^{10}$ cm$^{-3}$,
$N_H=10^{23}$ cm$^{-2}$, $\Gamma = \delta =15$ and, for simplicity, a
power-law electron energy distribution, extending from $\gamma _{\rm
min}=1$ to $\gamma _{\rm max}=10^7$ and with slope $n=3$ (thus
producing a spectrum with index $\alpha =1$) and $n=3.5$ ($\alpha
=1.25$). Solid lines refer to the case of the exact Klein-Nishina
cross section, while the dashed lines report the spectra calculated
with the ``step'' approximation. These spectra display a sudden drop
of the emission above $\nu _{\rm KN}$ (dotted vertical line,
calculated with $\nu _o=2.5\times 10^{15}$ Hz, the frequency of the
Ly$_{\alpha }$ line) due to the transition into the KN regime. In the
exact spectra the transition is smoother.}
\end{figure}

\noindent
Note that typically $\delta = \Gamma $ and thus this expression does
not contain the bulk Lorentz factor of the jet. In this case the
frequency above which the EC spectrum starts to drop due to the KN
effects only depends on the frequency of the soft photons, $\nu
_o$. 

Even if peaked at $\nu _o$, the external photon field is not
monochromatic, thus, instead of an abrupt break, the EC spectrum
shows a progressive decrease of the flux. Moreover, also the use of the
KN cross section contribute to smooth the spectrum. An example of the
high-energy spectrum (calculated for the case $n=10^{10}$ cm$^{-3}$ and
$N_H=10^{23}$ cm$^{-2}$) is shown in Fig. 8, assuming a simple
power-law electron energy distribution and two different spectral
indices, $\alpha =1$ and $\alpha =1.25$ (solid lines). For comparison
we also report (dashed lines) the same spectra calculated using the
``step function'' approximation for the KN cross section
(e.g. Ghisellini et al. 1998). In this case a sudden drop of the
emission, located at the frequency provided by Eq. (\ref{nukn}), is
visible. In the spectra calculated with the exact cross section this
feature is instead ``smoothed'' and the spectrum displays a continuous
decrease.

\subsection{Application: GB B1428+4217}

As an application of the model discussed above, in Fig. 9 we show the
comparison between the emission calculated with the EC model and the
{\it XMM-Newton} spectrum of GB B1428+4217, a powerful blazars located
at high redshift ($z=4.72$). The X-ray spectrum shows a large deficit
of soft photons with respect to the extrapolation of the power-law
fitting the data above 1 keV. The lack of soft photons in this and
other high-$z$ radio-loud quasars has been often interpreted as the
signature of the presence of large amounts of absorbing gas in the
quasar environment (Worsley et al. 2004). However, a direct
alternative is that the low-energy cut-off is due to the intrinsic
hardening of the EC emission below $\nu _{\rm br}$ (e.g. Celotti et
al. 2007, Tavecchio et al. 2007). Here we tried to reproduce the
spectrum with the {\it EC/BLR} model, trying to find the (probably not
unique) set of parameters which closely reproduces the data. The
corresponding {\it EC/BB} spectrum is also shown for comparison. We
fix $L_{\rm ill}=3\times 10^{47}$ erg/s accordingly to that provided
by optical-UV data, assumed to trace the disk emission. As before,
$R_{\rm in}$ is fixed close to the average of the values given by Kaspi
et al. and the Bentz et al. relations.  All the other relevant
parameters have been fixed to ``standard'' values, $n=10^{11}$
cm$^{-2}$, $N_H=10^{23}$ cm$^{-2}$, $\Gamma=12.3$, $n_1=2.2$, $\gamma
_{\rm min}=1$.

\begin{figure}
\centerline{\psfig{figure=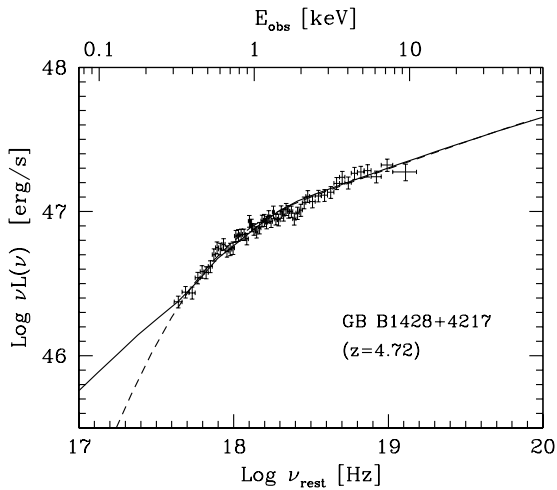,angle=0,width=8.5cm}}
\caption{{\it XMM-Newton} spectrum of GB B1428+4217 (Worsley et
al. 2004, 2006) (black crosses). The frequency is measured in the
quasar rest frame, while the observed energy is reported by the axis
on the top. The red solid line is the {\it EC/BLR} model reproducing
it, using the following parameters: $L_{\rm ill}=10^{47}$ erg/s,
$R_{\rm BLR}=3\times 10^{18}$ cm, $n=10^{11}$ cm$^{-2}$, $N_H=10^{23}$
cm$^{-2}$, $\Gamma=12.3$, $n_1=2.2$, $\gamma _{\rm min}=1$. The dashed
line shows the corresponding {\it EC/BB} spectrum.}
\end{figure}


As expected, the {\it EC/BLR} spectrum shows a tail below $3\times
10^{17}$ Hz (in the quasars rest frame) softer than $\nu^2$. As
discussed above, this seems to be an unavoidable and robust feature of
the {\it EC/BLR} spectra, due to the contribution of the BLR continuum
to the EC emission. Instead, if the lack of soft photons is the result
of absorption, the emission should continue to decrease almost
exponentially. Therefore, observations extending down to very low
energy and/or sources located at minor distance (so that the spectrum
is less redshifted) could be used to to discriminate between the
alternative interpretation for the deficit of low-energy photons.

\section{Discussion}

We have calculated the expected diffuse emission filling the BLR
region of radio-loud quasars, under quite general assumptions on the
physical parameters of the system. This external radiation field is
thought to dominate in the jet over the synchrotron photons (External
Compton emission), hence contributing to the bulk of the inverse
Compton emission (Sikora et al. 1994). In all cases, few emission
lines dominate the total luminosity. However, for frequencies below
$10^{15}$ Hz, the emission is dominated by the continuum (mainly from
free-free emission). We have presented for the first time the EC
spectra expected from the comptonization of realistic BLR spectra,
showing that the black body approximation (peaked at 1.5 $\nu
_{Ly _{\alpha}}$) for the external radiation field is rather good down to
the soft X-ray band. Below 1 keV, however, the spectra calculated with
the BLR spectra show an excess, due to the important contribution of
the BLR continuum.

The presence of this tail in the EC emission can be used to
effectively discriminate the origin of the observed lack of soft
photons observed in several high-redshift radio-loud quasars. This
evidence is generally attributed to the absorption by a huge amount
(column densities of the order of $10^{22}$ cm$^{-2}$) of gas in the
vicinity of the jet (e.g., Elvis et al. 1994, Cappi et al. 1997,
Reeves et al. 1997, Fiore et al. 1998, Fabian et al. 2001a,b, Bassett
et al. 2004, Yuan et al. 2005, Page et al. 2005), whose presence is
predicted to be important especially at high redshifts (e.g. Fabian et
al. 1999). However, the fact that, apparently, only the radio-loud
quasars display such a deficit of soft photons can suggest that it can
be due to the intrinsic shape of the EC emission dominating in this
energy band (e.g. Tavecchio et al. 2007, Sambruna et
al. 2007). Indeed, as we have shown for the specific case of GB
B1428+4217, this possibility seems to be viable even considering a
realistic spectrum for the BLR emission. However, after the hardening
around 1 keV, the EC spectrum should display a softer tail extending
at low energies. Future observations could be effectively reveal such
a tail. On the contrary, very hard spectra extending down into the
soft X-ray band can be a strong support to the absorption scenario.

Another interesting point related to the physics of jets concerns the
value of the minimum Lorentz factor of the emitting electrons, $\gamma
_{\rm min}$. Having fixed the spectral shape of the seed photons for
the EC process, the position of the break in the spectrum is only
related to the product $\Gamma ^2\gamma _{\rm min}^2$ (see also
Maraschi \& Tavecchio 2003, Ghisellini et al. 2007). Values of $\gamma
_{\rm min}$ larger than 1 would move the position of the break at
frequencies larger than observed. The break could be located at the
right position decreasing the value of the bulk Lorentz factors.  For
instance assuming $\gamma _{\rm min}=2$ in the case of GB B1428+4217
would require $\Gamma =6$. These values however, seem to be
inconsistent with those inferred from the modelling of the SED,
usually in the range 15-20 (e.g. Celotti \& Ghisellini 2008).

The discussion above does not consider the contribution to the EC
process of the soft radiation coming from other regions, as the disk
itself (directly or reprocessed) or from the putative obscuring torus
(B{\l}a{\.z}ejowski et al. 2000). In the cases the corresponding EC
emissions should extend at very low frequencies (close to the optical
band), unless the value of $\gamma _{\rm min}$ is rather large
(e.g. Celotti et al. 2007). Therefore in this case the absorption
would be the only possible explanation for the presence of cut-offs
the soft X-ray band. Moreover, we also neglect in the discussion the
contribution of the SSC component, important especially in the soft
X-ray band. However, from the observational point of view the the SSC
component seems to be less important in powerful sources (maybe due to
an increasing importance of the accretion disk and then of the
external radiation field: e.g., Celotti et al. 2007), leaving an
almost ``naked'' EC component.

Besides the problems related to the description of the soft EC
spectrum, the modelling of the BLR radiation field is rather important
also in view of the estimate of the opacity of the quasar environment
to $\gamma$-rays. Indeed absorption of multi GeV photons inside the
quasar can represent a serious problem for the possible use of blazars
as probes of the optical-UV cosmic background. Recent works (Liu \&
Bai 2006, Reimer 2007) uses approximate spectra of the BLR to
calculate the expected opacity to gamma-rays. We note here that
another effect, besides internal absorption, is important in shaping
the intrinsic high-energy spectrum of powerful quasars. Indeed, as we
have discussed in Sect. 3.2.2, the EC spectrum above few tens of GeV
is affected by a strong depression, due to KN effects. The coupling of
these effects (KN regime and internal absorption) will probably result
in a rather soft and dim emission above $\approx 10-20$ GeV, making it
rather difficult to use blazars as probes of the optical-UV
extragalactic background, unless some other radiative mechanisms
dominates the high energy emission.

\section*{Acknowledgments}
We thank Gary Ferland for maintaining his freely distributed code
CLOUDY.

\end{document}